\documentclass[aps,longbibliography,prd,twocolumn,groupedaddress]{revtex4}
\usepackage{graphicx}
\usepackage{amsmath,amssymb,amsfonts}
\usepackage[colorlinks,citecolor=blue,linkcolor=blue,urlcolor=blue]{hyperref}

\usepackage[english]{babel}

\usepackage[autostyle, english = british]{csquotes}
\MakeOuterQuote{"}

\begin{document}

\title{Interior metric and ray-tracing map in the firework  black-to-white hole transition}

\author{C. Rovelli, P. Martin-Dussaud}
\email{pmd@cpt.univ-mrs.fr}
\affiliation{Aix Marseille Univ, Universit\'e de Toulon, CNRS, CPT, Marseille, France}

\date{\small\today}

\begin{abstract}
\noindent
The possibility that a black hole could tunnel into a white hole has recently received  attention. Here we present a metric that improves the "firework" geometry: it describes this entire process and solves the Einstein equations everywhere except on a small transition surface that corresponds to the quantum tunnelling. We also explicitly compute the corresponding ray-tracing map from past infinity to future infinity.
\end{abstract}

\maketitle 

\section{Fireworks}

Black holes have become common astrophysical objects and all recent observations strengthen the conclusion that the most accurate theory we have to describe them is still centenarian general relativity. Yet classical general relativity leaves questions open. What happens ultimately to the in-falling matter? Is information lost after Hawking evaporation? There is no consensual answer to these questions yet. 
A scenario to address them has recently raised interest: the possibility of a quantum tunnelling from a black hole to a white hole \cite{Rovelli2014,Haggard2014,DeLorenzo2016,Christodoulou2016,Christodoulou2018,DAmbrosio2018,Bianchi2018}. This transition is allowed by general relativity, provided that quantum theory permits the violation of Einstein's equation (by a tunnelling process) in a small compact spacetime region. A spacetime realising this scenario, called the \textit{firework} metric because matter inside the hole can explode out of the white hole after the tunnelling, was found in  \cite{Haggard2014}.  The calculation of the quantum probability for the process has been addressed in \cite{Christodoulou2016,Christodoulou2018}, using covariant loop quantum gravity; the effect of the Hawking evaporation and the relevance of the scenario for the information loss paradox have been recently discussed in \cite{Bianchi2018}. A number of possible phenomenological consequences of this process have been studied in \cite{Barrau2014d,Barrau2014b,Barrau2016a,Haggard2016,Barrau2017,Rovelli2018b,Rovelli2018c}. 

Here we present an improvement on the firework metric discovered in \cite{Haggard2014}. Following \cite{Bianchi2018} we distinguish between two physically distinct quantum phenomena relevant in this process. In the terminology of  \cite{Bianchi2018}, region $A$ is the Planckian-curvature region around the singularity where the interior black hole metric continues to a white hole metric.   Physically, this describes the interior bounce, a stage called "Planck star" \cite{Rovelli2014}. This transition can be modelled by a smooth joining of two Kruskal spacetimes, a possibility noted by several authors \cite{Synge1950,Peeters:1994jz} and recently discussed in \cite{DAmbrosio2018}. The proper quantum tunnelling is then confined to a small region $B$ \cite{Bianchi2018}, which surrounds the end of the apparent horizon of the black hole. Here we give a metric that satisfies the Einstein equations (in the sense of \cite{DAmbrosio2018}) everywhere except in this small region.  Incidentally, we cure a pathology of the original firework metric: a conical singularity at the cusp point of the quantum region. 

Specifically, we present a metric that has the following properties. 
\begin{enumerate}\itemsep=0mm
\item[(i)] It describes the fall and collapse of a thin null spherical shell of matter, which bounces at a minimal radius \emph{inside its Schwarzschild radius}, and then expands forever. (This scenario is of course not allowed by the \emph{classical} theory.)
\item[(ii)] The metric satisfies Einstein equations almost everywhere. Due to Bhirkhoff's theorem, the shell's interior is therefore a portion of Minkowski spacetime, while the exterior is almost everywhere locally isomorphic to a portion of Kruskal spacetime.
\item[(iii)] We neglect the thickness of the shell. 
\item[(iv)] The spacetime is spherically symmetric. As a consequence, the spacetime can be represented pictorially by a Penrose diagram.
\item[(v)] We assume that the process is invariant under time-reversal. In particular, we disregard the dissipative effects such as the Hawking radiation. The extension to non time-reversal metrics will be studied elsewhere. 
\item[(vi)] The timelike and null geodesics are continuous through the $r=0$ singularity.
\end{enumerate}

\section{Kruskal origami}

\begin{figure}[b]
	\includegraphics[width = .9 \columnwidth]{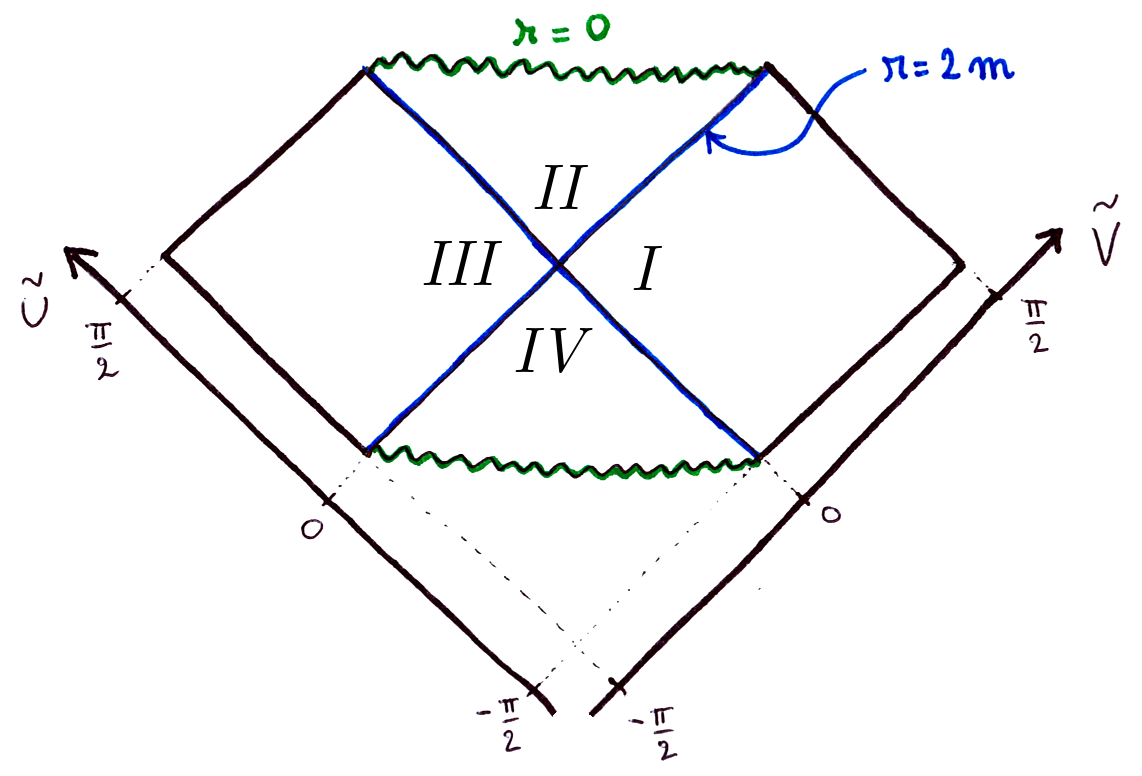}
	\caption{Penrose diagram of the Kruskal spacetime. $\tilde {U}$ and $\tilde {V}$ are the (Cartesian) Penrose coordinates.}
	\label{Kruskal}
\end{figure}

\begin{figure}[t]
	\centering
	\includegraphics[width = .7 \columnwidth]{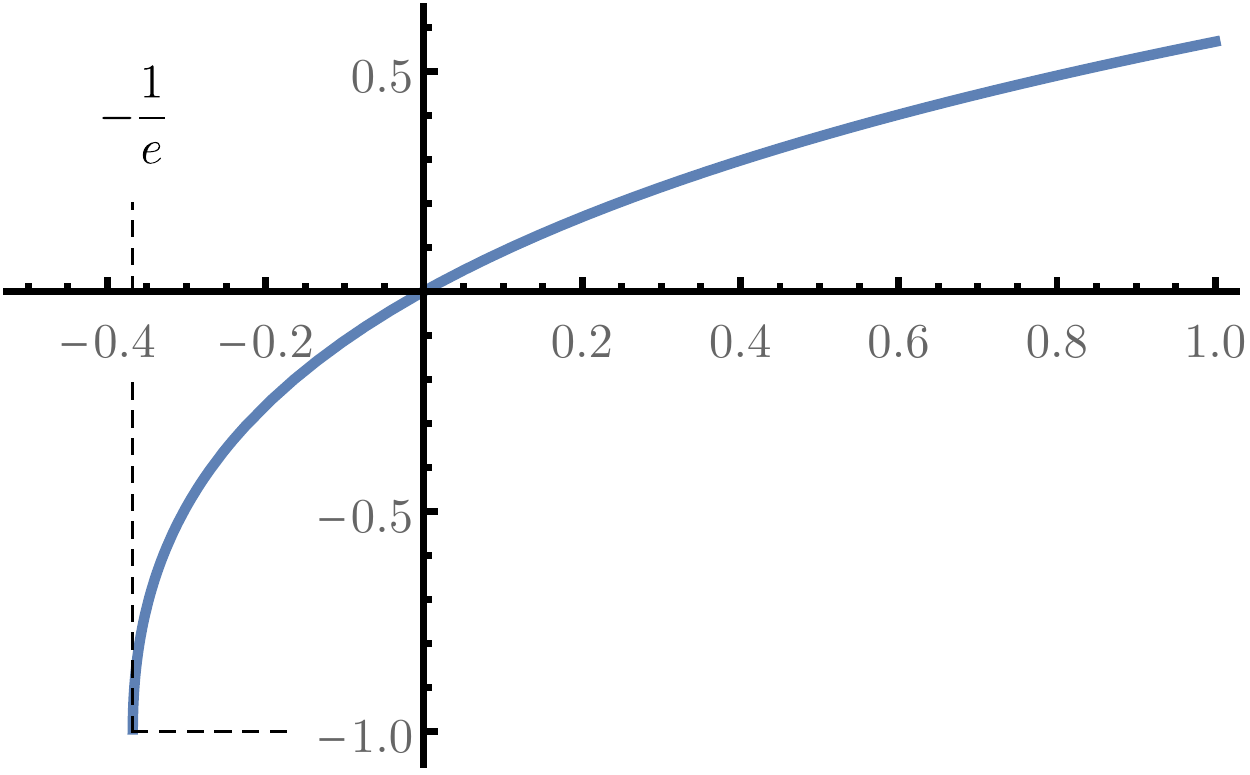}
	\caption{Graph of the upper branch of the Lambert $W$ function.}
	\label{Lambert}
\end{figure}

\paragraph{The Kruskal spacetime.}
The maximal extension of the Schwarzschild black hole is the Kruskal spacetime. Its Penrose diagram is recalled in Figure \ref{Kruskal}. The metric is given, in terms of the Kruskal coordinates $(U,V)$, by 
\begin{equation}\label{Kruskal metric}
{\rm ds}^2 = - 32 m^3\; \frac{e^{-r/2m}}{r} \, dU \, dV + r^2 d\Omega^2, 
\end{equation}
where $d\Omega^2 = d\theta^2 + \sin^2 \theta \, d \phi^2$ is the metric of the unit sphere and $r$ is the function defined by
\begin{equation}\label{Kruskal radius}
r(U,V) = 2m \left[1+ W\left( -\frac{UV}{e} \right) \right]. 
\end{equation}
The function $W$ is the upper branch of the {Lambert W function}. It is a increasing function defined by the equation $x = W(x) e^{W(x)}$ and its graph is shown in Figure \ref{Lambert}.

The Kruskal coordinates $(U,V)$ are expressed in terms of the Penrose coordinates $(\tilde{U},\tilde{V})$ by the relations
\begin{equation}\label{Kruskal coordinates}
\left\{ \begin{array}{l}
U = \tan \tilde {U} ,\\
V = \tan \tilde {V} .
\end{array} \right.
\end{equation}
The coordinates $\tilde{U}$ and $\tilde{V}$ are Cartesian coordinates for the diagram of the Figure \ref{Kruskal}. Finally, in the region $I$, the null-coordinates $(u,v)$ are expressed in terms of the Kruskal coordinates by the relations :
\begin{equation}\label{null-coordinates}
\left\{ \begin{array}{l}
u = - 4 m \log ( - U ) , \\
v = 4m \log V .
\end{array} \right.
\end{equation}

\vspace{\baselineskip}

\begin{figure}[h!]
	\includegraphics[width = .9 \columnwidth]{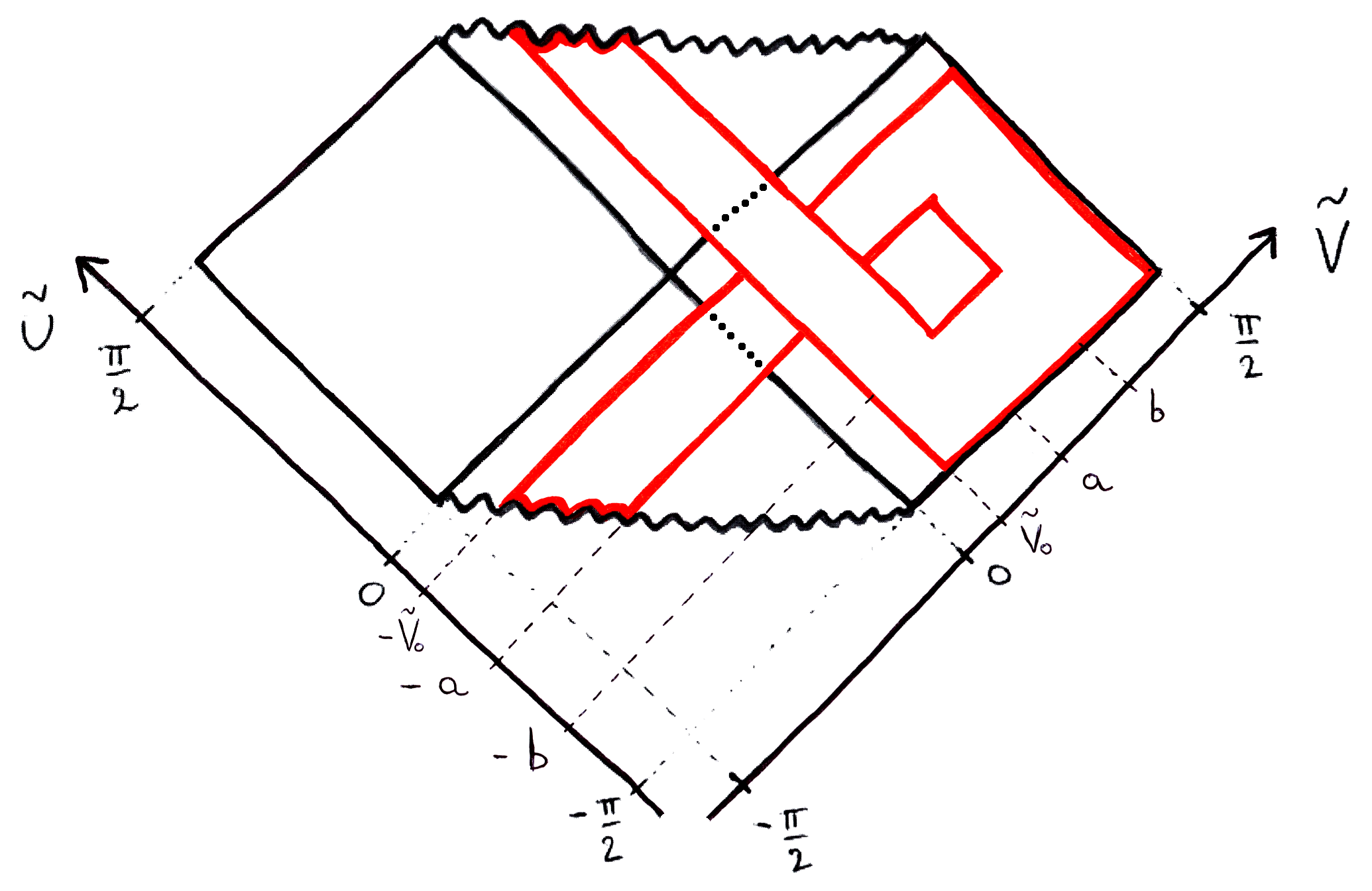}
	\caption{Penrose diagram of the Kruskal spacetime. The red straight lines are null, and the two red wavy lines will be identified after "squashing the arms". The inside region thus delimited is the spacetime of interest for us.}
	\label{cut}
\end{figure}

\paragraph{A snip of the scissors.}
We now consider the portion of Kruskal spacetime marked out by the red line in Figure \ref{cut}. It is a  connected region consisting of two "arms", one touching the past singularity, the other the future one. You may notice a local double covering (where the two arms cross), which raises no peculiar difficulty. 

\vspace{\baselineskip}

\begin{figure}[h!]
	\includegraphics[width = .8 \columnwidth]{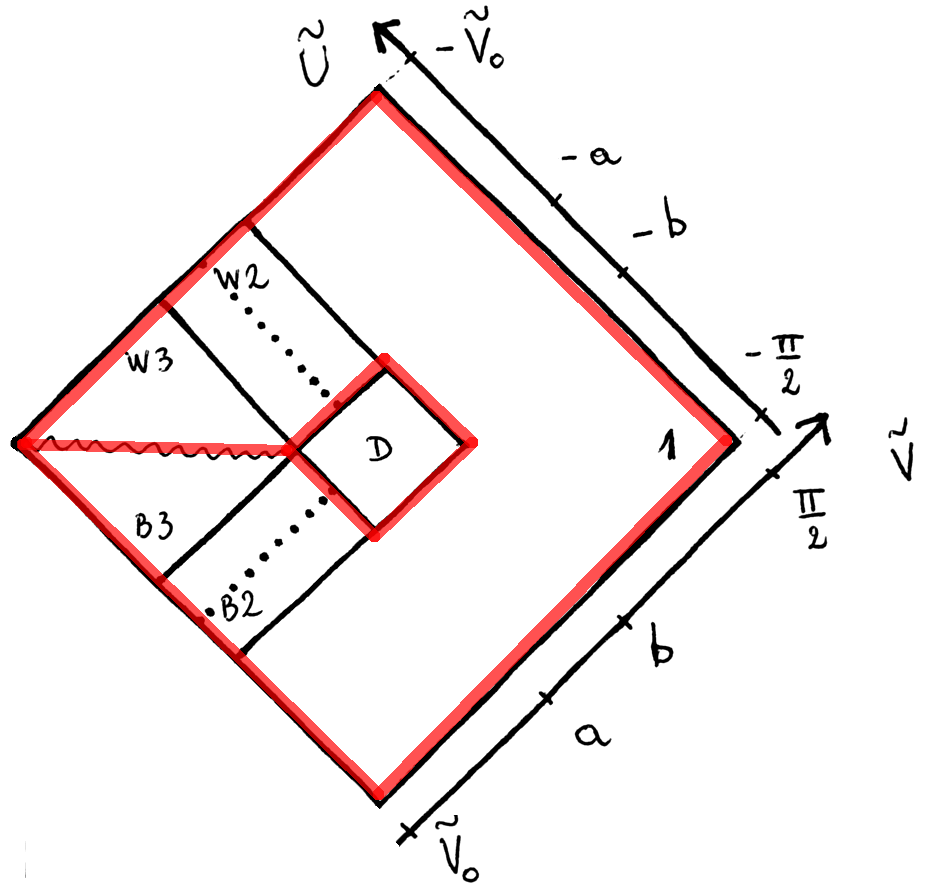}
	\caption{Penrose diagram of the outside of the null shell. The dotted lines are the two horizons at $r=2m$.}
	\label{origami}
\end{figure}

\paragraph{Squashing the arms.}
The modelling of the black-to-white hole transition is achieved through the identification between the past and the future singularity. Heuristically, it consists in "squashing the arms until the hands match". The Penrose diagram of the resulting spacetime is represented in Figure \ref{origami}. 

The expression of the metric is still given by equations \eqref{Kruskal metric} and \eqref{Kruskal radius}, where the Kruskal coordinates $(U,V)$ are given in terms of the Penrose coordinates $(\tilde{U},\tilde{V})$ by
\begin{align} \label{Origami1}
\text{[lower half]} & \left\{ \begin{array}{l}
U = \tan f_B(\tilde  U) \\
V = \tan \tilde {V} 
\end{array} \right. \\ \label{Origami2}
\text{[upper half]} &\left\{ \begin{array}{l}
U = \tan \tilde {U}  \\
V = \tan f_W(\tilde {V})
\end{array} \right.
\end{align}
where the two functions $f_B$ and $f_W$ are differentiable and defined piecewise such that
\begin{align} \label{f_B}
f_B( \tilde{U}) = \left\{ \begin{array}{llr}
 \tilde{U} & {\rm for }& \tilde{U} \in [- \frac{\pi}2 , -b]\\
 f_B(\tilde{U}) & {\rm for} & \tilde{U} \in [- b , -a] \\
\tilde{U} + \frac{\pi}{2}  & {\rm for  }&  \tilde{U} \in [-a,-\tilde{V}_0]
\end{array} \right.
\end{align}
and
\begin{align} \label{f_W}
f_W( \tilde{V}) = \left\{\begin{array}{llr}
 \tilde{V} - \frac{\pi}{2} & {\rm for }& \tilde{V} \in [\tilde{V}_0,a] \\
 f_W(\tilde{V}) & {\rm for} & \tilde{V} \in [a , b] \\
\tilde{V} & {\rm for  }&  \tilde{V} \in [b,\frac{\pi}{2}].
\end{array} 
\right. 
\end{align}
For the intermediate intervals ($[- b , -a]$ for $f_B$ and $[a , b]$ for $f_W$) one can choose any continuous and monotonous function which joins "smoothly enough" with the other pieces. 

\begin{figure}[h!]
	\includegraphics[width = .8 \columnwidth]{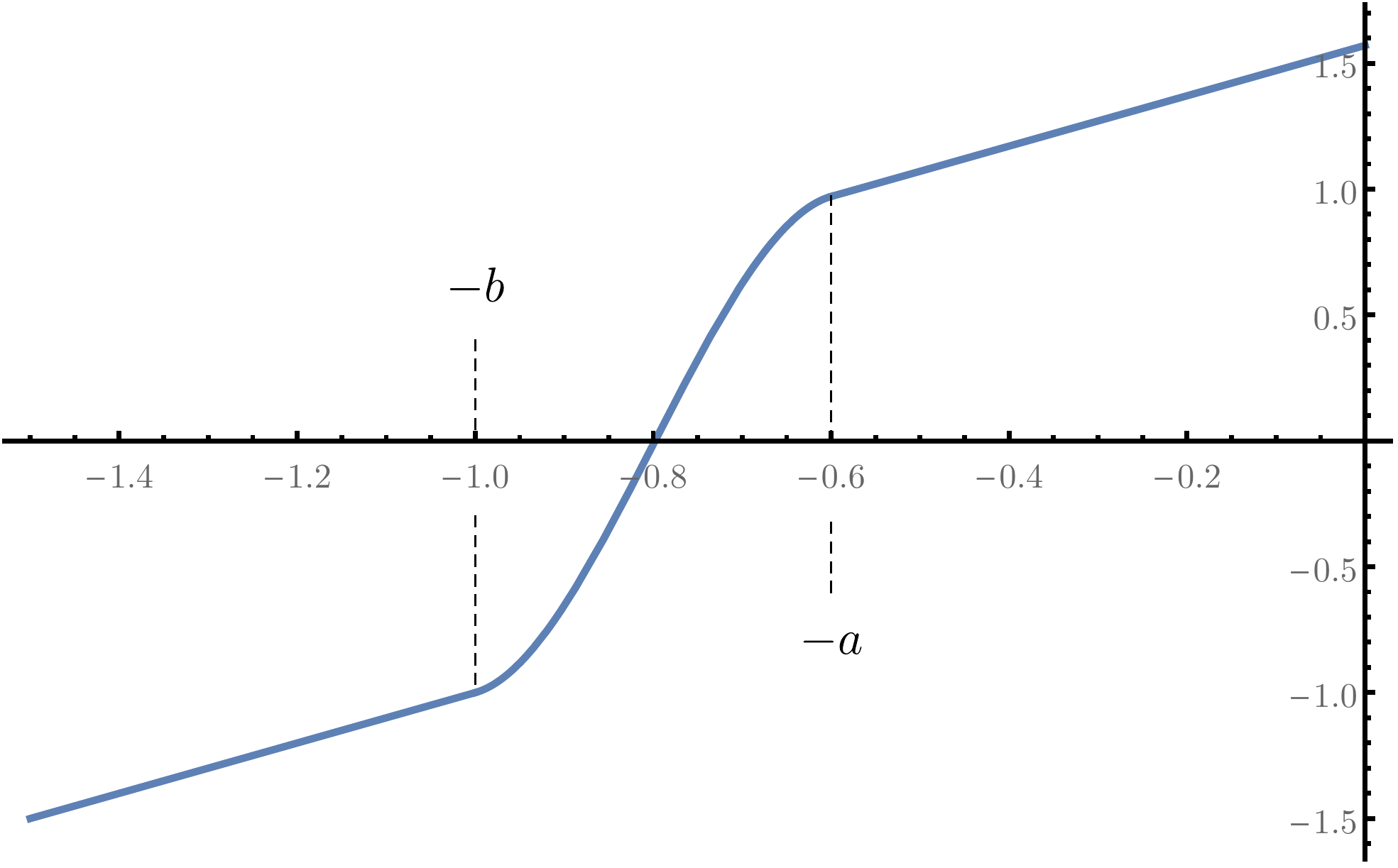}
	\caption{Graph of the function $f_B$. On the interval $[-b,-a]$ it is a polynomial of degree $3$. It is linear elsewhere. Here we have chosen $a=0.6$ and $b=1$.}
	\label{f_Bgraph}
\end{figure}

The minimal smoothness required is $\mathcal{C}^1$. Indeed, the two junction conditions for null hypersurfaces have to be satisfied along the null geodesics $\tilde{V}=a$, $\tilde{V}=b$, $\tilde{U}=-a$ and $\tilde{U}=-b$. The first condition is the continuity of the induced metric on the hypersurface. This requires the continuity of the functions $f_B$ and $f_W$. The second condition is the continuity of the extrinsic curvature, which imposes the continuity their derivatives. In the following, we will choose, in the intermediate interval, a polynomial of degree 3 which is sufficient for $f_B$ or $f_W$ to be $\mathcal{C}^1$ (see Figure \ref{f_Bgraph}).

\vspace{\baselineskip}

\paragraph{Across the singularity.}

\begin{figure}[h!]
	\includegraphics[width = 0.5 \columnwidth]{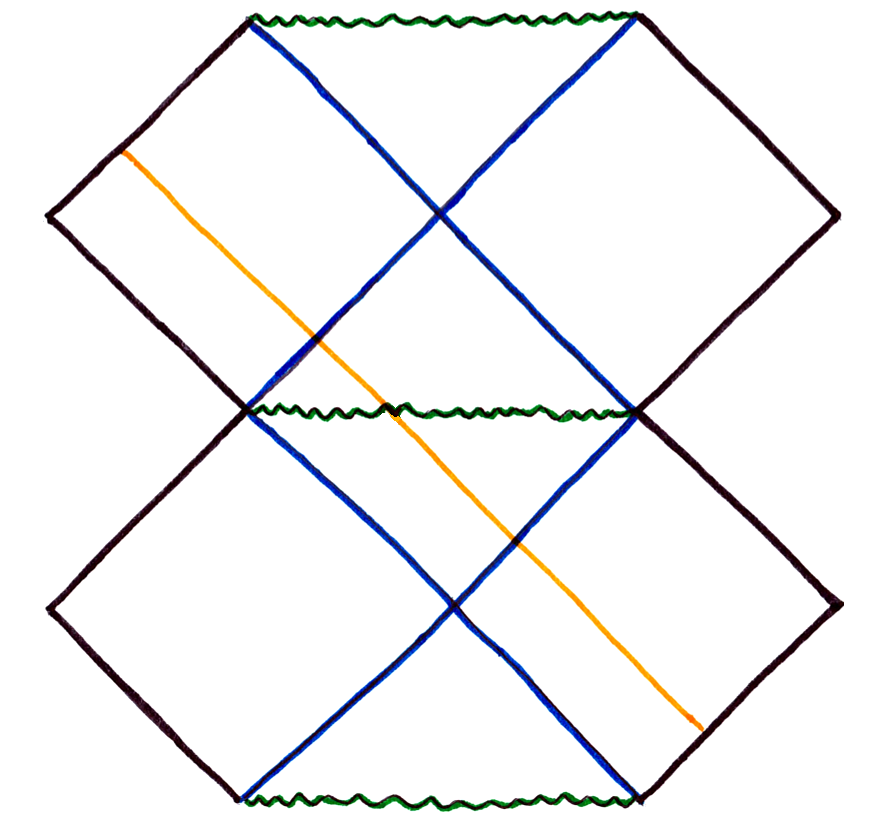}
	\caption{Penrose diagram of the two Kruskal spacetimes joined at the singularity. The orange line represents an ingoing null geodesic crossing the singularity.}
	\label{2Kruskal}
\end{figure}

The regions $B3$ and $W3$ touch along the singularity. There is no difficulty here. It has been repeatedly noticed \cite{Synge1950,Peeters:1994jz} that it is possible to match the future singularity of a Kruskal diagram to the past singularity of another (see Figure \ref{2Kruskal}). The metric is singular there, but there is a natural prescription for the geodesics to go across the singularity, requiring conservation of linear and angular momentum \cite{Peeters:1994jz}. As argued in \cite{DAmbrosio2018}, the resulting spacetime can be seen as the $\hbar\to 0$ limit of the effective metric of a non singular spacetime where quantum gravity bounds curvature. This effective metric would be given by 
\begin{equation}
ds^2 = - \frac{4(\tau^2 + l)^2}{2m - \tau^2} d\tau^2 + \frac{2m-\tau^2}{\tau^2+l} dx^2 + (\tau^2 + l)^2 d\Omega^2
\end{equation}
with a constant $l \sim (m\hbar)^{1/3}$. When $\hbar \to 0$, $l \to 0$, and the the metric boils down to the usual Schwarzschild metric inside the black-hole with the usual Schwarzschild time $t$ and radius $r$ given by $t = x$ and $r = \tau^2$. So there is a sense in which the two glued Kruskal spacetimes are still a solution of Einstein's equations. We take this as a simplified model of the quantum transition across the singularity (region $A$ in the terminology of \cite{Bianchi2018}). 

\vspace{\baselineskip}

Finally, the metric is well defined all around the central diamond $D$. This metric is Ricci-flat everywhere (vacuum solution), up to the $r=0$ surface that separates $B3$ and $W3$ where it still solves the Einstein's equations in the sense of \cite{DAmbrosio2018}.

\vspace{\baselineskip}

\paragraph{The diamond $D$.}
The central diamond $D$ is the quantum tunnelling region (region $B$ in the terminology of \cite{Bianchi2018}).
The simplest possibility to define a metric in this region is to simply extend the metric of $B2$ and of $W2$, respectively up to and down to the horizontal line $\tilde{U} + \tilde{V} = 0$. Then, the first junction condition along this hypersurface imposes :
\begin{equation}
f_W (x) = - f_B(- x). 
\end{equation}
However, the second junction condition can never be satisfied, because otherwise it would define an exact solution of Einstein's equations with the same past but a different future as a standard collapse metric, which has an event horizon.  The discontinuity of the extrinsic curvature therefore encodes the quantum transition in this region, as studied in \cite{Christodoulou2016,Christodoulou2018}. The novelty is that now this tunnelling region is confined within the diamond. The discontinuity in the extrinsic curvature could be smoothed out by modifying the metric in a small neighbourhood of the discontinuity. This is possible at the price of violating Einstein's equations in this neighbourhood.

\section{Relighting the fireworks}

The metric constructed in the previous section describes the spacetime \emph{outside} the bouncing null shell. Inside the shell, spacetime is flat, therefore a portion of Minkoswki spacetime. What remains to be done is to glue a patch of Minkoswki along the collapsing and the emerging null shell. This is done in a similar way to the well-known model of Vaidya \cite{Vaidya1951}. 

\begin{figure}[h!] 
	\centering
	\includegraphics[width = 0.5 \columnwidth]{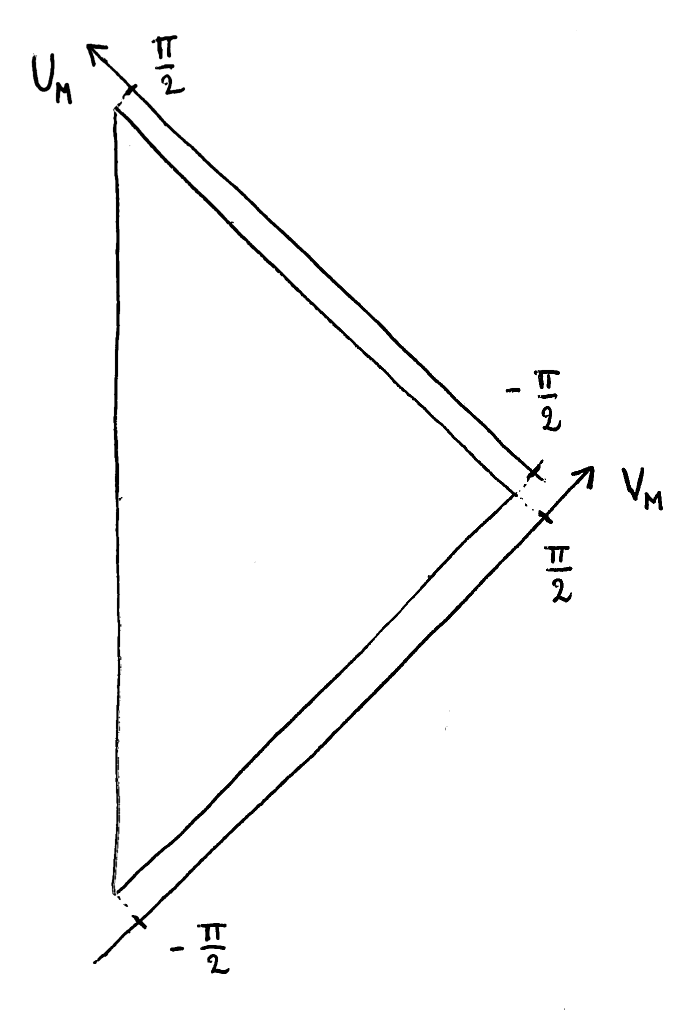}
	\caption{Penrose diagram of Minkowski spacetime.}\label{Minkowski diagram}
\end{figure}

The Minkowski metric in Penrose coordinates reads 
\begin{equation}\label{Minkowski metric}
{\rm ds}^2 = - \frac{dU_M dV_M}{\cos^2 U_M \cos^2 V_M} + r_M^2 d\Omega^2 ,
\end{equation}
with
\begin{equation}
r_M = \frac{1}{2} \left( \tan V_M - \tan U_M \right) . \label{rad}
\end{equation}
The Penrose diagram is shown in Figure \ref{Minkowski diagram}.
The null coordinates are given in terms of the Penrose coordinates by 
\begin{equation}\label{Minkoswki coordinates}
\left\{ \begin{array}{l}
u = \tan U_M ,\\
v = \tan V_M .
\end{array} \right.
\end{equation}

It is possible to glue a portion of Minkowski to the Kruskal origami by matching the value of the radius along a null ingoing geodesics ($V_M = constant$) for Minkowski with the value of the radius along the line $\tilde{V}=\tilde{V}_0$ of the Kruskal origami. 
This matching defines a map $U_M(\tilde U)$ given by 
\begin{equation}\label{BUM}
\tan U_M(\tilde  U)= v_0 -4m \left[ 1 + W \left( - e^{\frac{v_0}{4m}-1} \tan f_B(\tilde  U) \right) \right] ,
\end{equation}
with $v_0 \overset{\rm def}= 4m \log \tan \tilde{V}_0$.
Then the first junction condition is satisfied. The violation of the second is the effect of the stress-energy tensor of the collapsing shell. Finally, the same procedure can be applied for the outgoing null geodesics along the line $\tilde{U}=-\tilde{V}_0$, with the condition
\begin{equation} \label{WVM}
\tan V_M(\tilde{V}) = - v_0 + 4 m \left[ 1 + W \left( -  e^{\frac{v_0}{4m}-1} \tan f_B ( - \tilde{V} ) \right) \right].
\end{equation}
This completes the construction of the new spacetime for black-hole fireworks.

\vspace{\baselineskip}

\begin{figure}[h!]
	\centering
	\includegraphics[width = .6 \columnwidth]{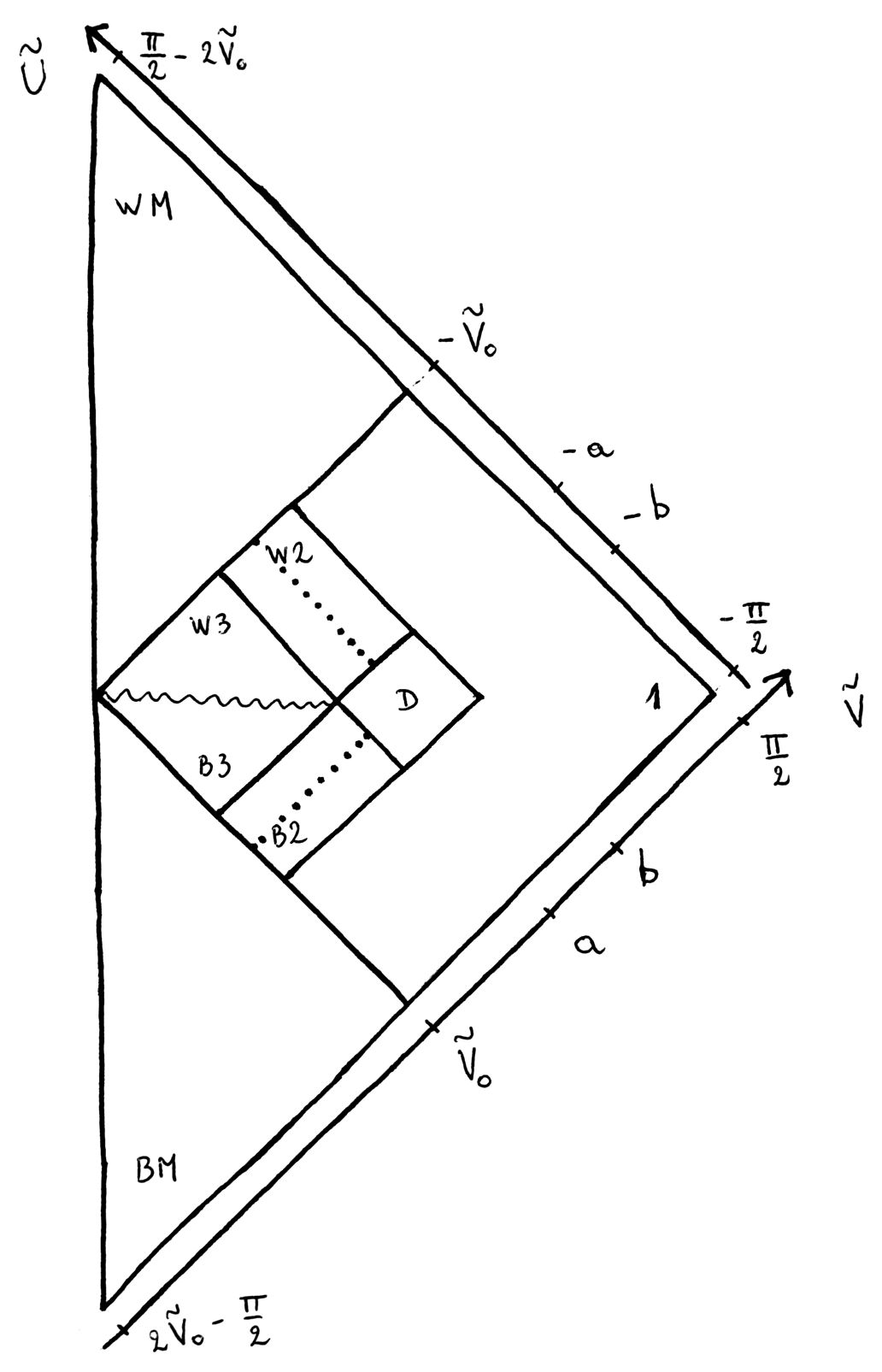}
	\caption{Penrose diagram of the new spacetime for fireworks.}
	\label{final1}
\end{figure}

\paragraph{Penrose diagram of the new spacetime.}
A Penrose diagram for the new spacetime has been drawn in Figure \ref{final1}.

To make an easy drawing, we have chosen to impose that the line $r_M=0$ should be straight and vertical, which is possible provided the map $V_M(\tilde{V})$ in $BM$ is given by
{\small
\begin{equation} \label{BVM}
\tan V_M(\tilde{V}) = v_0 - 4 m \left[ 1 + W \left( - e^{\frac{v_0}{4m}-1} \tan f_B ( \tilde{V} - 2 \tilde{V}_0 ) \right) \right] ,
\end{equation}
}
and the map $U_M(\tilde{U})$ in $WM$ is given by
{\small
\begin{equation} \label{WUM}
\tan U_M(\tilde  U)= - v_0 + 4m \left[ 1 + W \left( - e^{\frac{v_0}{4m}-1} \tan f_B(- \tilde  U - 2 \tilde{V}_0 )  \right) \right] .
\end{equation}
}

The metric outside the shell is Kruskal, described by equations \eqref{Kruskal metric}, \eqref{Kruskal radius}, \eqref{Origami1}, \eqref{Origami2}, \eqref{f_B} and \eqref{f_W}. The metric in the two regions $BM$ and $WM$ is Minkowski, given by equations \eqref{Minkowski metric}, \eqref{rad}, and respectively, \eqref{BUM} and \eqref{BVM} for $BM$, and \eqref{WUM} and \eqref{WVM} for $WM$.

\vspace{\baselineskip}

\begin{figure}[b]
	\centering
	\includegraphics[width = .6 \columnwidth]{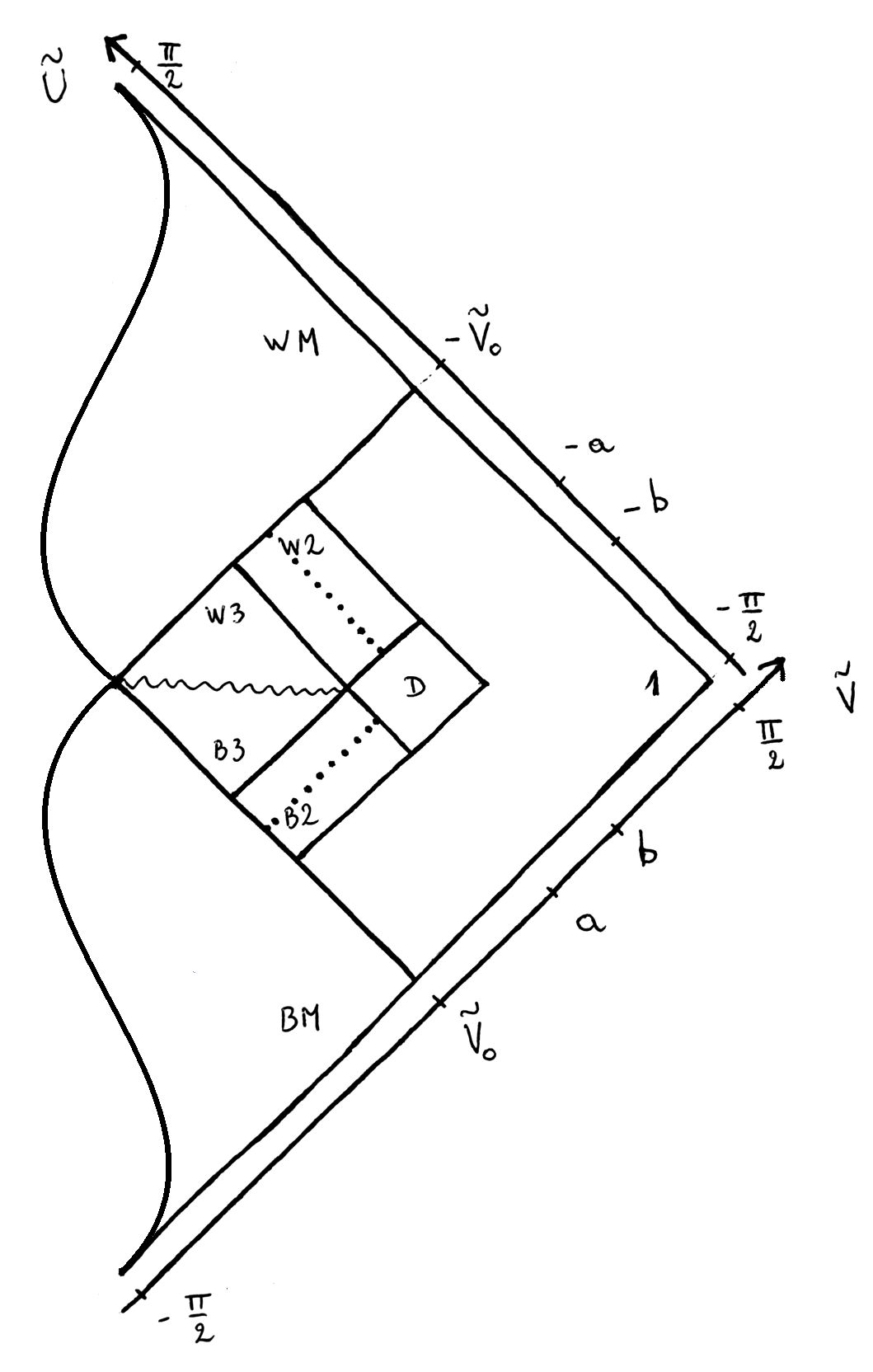}
	\caption{Another possible Penrose diagram.}
	\label{final2}
\end{figure}

\paragraph{Another Penrose diagram.}
Another way to proceed would be to impose 
\begin{equation}
\left\{
\begin{array}{ll}
V_M(\tilde{V}) = \tilde{V} & \text{in } BM \\
U_M(\tilde{U}) = \tilde{U} & \text{in } WM
\end{array}
\right.
\end{equation} 
and then, to draw the Penrose diagram accordingly (see Figure \ref{final2}). The only difference is the shape of the line $r_M = 0$, which is now given by 
\begin{equation}
\tan \tilde{V} = v_0 -4m \left[ 1 + W \left( - e^{\frac{v_0}{4m}-1} \tan f_B(\tilde  U) \right) \right] 
\end{equation}
in the region $BM$ and 
\begin{equation}
\tan \tilde{U} = - v_0 + 4 m \left[ 1 + W \left( -  e^{\frac{v_0}{4m}-1} \tan f_B ( - \tilde{V} ) \right) \right]
\end{equation}
in the region $WM$.

\section{The ray-tracing map}

In reference \cite{Bianchi:2014bma}, it is shown that the energy flux of the Hawking radiation of a quantum field on a given spherically symmetric metric, and its entropy, can be computed, under some approximations, once given the so-called ray-tracing map.  The ray-tracing map is the map from past null infinity to future null infinity defined by the radial null geodesics. Here we compute explicitly the ray-tracing map of the metric we have defined.  The resulting properties of Hawking energy and entropy will be  studied elsewhere.   

In the construction above, the ray-tracing-map is directly given by the equation of the line $r_M=0$ written in the null-coordinates. We show the following expression:

\begin{widetext}

\begin{equation}
u(v) = \left\{
\begin{array}{ll}
- 4m \log \left[ - \tan f_B^{-1} \left( \arctan \left[ \left( 1- \frac{v_0 - v}{4m} \right) e^{-v/4m} \right] \right)  \right] & \text{if } v \leq v_0 , \\
& \\
- v_0 + 4m \left[ 1 + W \left( - \tan f_B ( - \arctan  e^{v/4m} ) e^{\frac{v_0}{4m}-1}  \right) \right] & \text{if } v_0 < v.
\end{array} \right.
\end{equation}
\end{widetext}

One can check that it is continuous for $v=v_0$ with
\begin{equation}
u(v_0)  = - v_0.
\end{equation}

Usually, the ray-tracing map is defined such that $u(0) = 0$, which is not the case here. It could be easily obtained by addition of a constant. 

\begin{figure}[h!]
	\centering
	\includegraphics[width = 1 \columnwidth]{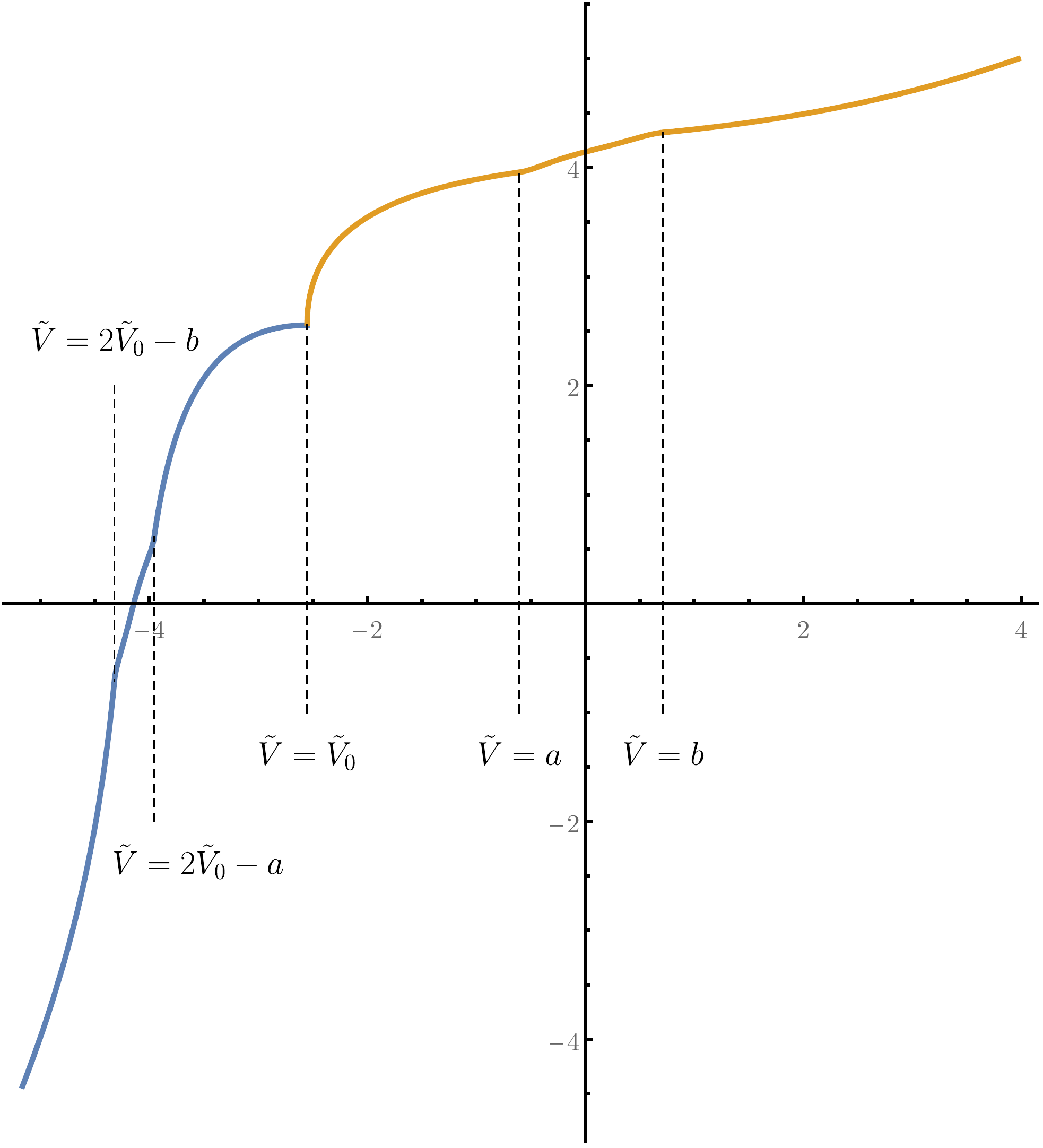}
	\caption{Graph of the ray-tracing map $u(v)$. Here we have chosen the parameters $m=0.4$, $\tilde{V}_0 = 0.2$, $a=0.6$ and $b=1$.}
	\label{plot}
\end{figure}

The ray-tracing map is plotted in Figure \ref{plot}, for the choice of $f_B$ plotted in Figure \ref{f_Bgraph}. 

Unsurprisingly the ray-tracing is symmetric with respect to the line $u=-v$, which was expected from the time-reversal symmetry of the constructed spacetime. Moreover, in the limit $v \to \pm \infty$, the ray-tracing map behaves like
\begin{equation}
u(v) \sim v - 4m \log \frac{|v|}{4m},
\end{equation}
which is also expected from the usual Vaidya spacetime. 

The continuity of the ray-tracing map over all the range of $v$ is an interesting novelty of the metric proposed here compared to the former firework metric,  where the ray-tracing map was incomplete around the singularity \cite{Bianchi:2014bma}. It would be the task of future works to study carefully the radiative aspects of this spacetime.

\section{Conclusion and open questions}

We have constructed a complete metric for a spacetime describing the tunnelling of a black-hole into a white-hole. It satisfies Einstein equation everywhere except in the small quantum region called the "diamond". Compared to the previous Haggard-Rovelli construction, it introduces several improvements such as the continuity of the transition through the singularity.  Several questions remain open. 

Firstly, white holes are unstable with respect to small perturbations of the ambient matter (classical instability) and also with respect to particle creation (quantum instability). In \cite{DeLorenzo2016} it was pointed out that both  classical and  quantum instabilities may affect the Haggard-Rovelli firework metric. The same analysis could be carried out here, similarly leading to the conclusion that the new metric is still unstable. The same article \cite{DeLorenzo2016} was also suggesting a way to make the metric stable by dropping time-reversal symmetry. This is one reason for extending this work to the non time-symmetric case. 

More recently, the idea was suggested that taking the Hawking evaporation into account can naturally lead to the stability of the metric \cite{Bianchi2018}. In this framework, the black-to-white hole transition occur when the black hole reaches Planckian size, after it has evaporated for a time of about $m^3 / \hbar$. Then the instability problem is naturally cured since there is no possible sub-planckian perturbations. The resulting spacetime is strongly asymmetric, and the white hole would appear to be a remnant. The explicit construction of an asymmetric spacetime that would account for the remnant status of white hole is still to be done, but surely we would get inspiration from the symmetric case just described.  An extensive detailed discussion about the black-to-white hole stability as well as a discussion about some of the observational aspects of white holes can be found in reference \cite{Rovelli2018c}. The question of the stability of white holes is not closed, and many issues, such as the white hole radiation (see  for instance \cite{Hsu2012}) remain unaddressed.

The metric given here should help study the black-to-white hole transition phenomenon. A convincing treatment, however, must include a full quantum calculation. In particular, the classical construction does not allow us to say anything about the tunnelling rate $p$ of the transition, and therefore the black hole lifetime as a function of the mass.   Naive Euclidean or Lorentzian tunnelling probabilities estimations are hard in this context. Preliminary estimates relying on Loop Quantum Gravity have given a tunnelling probability
\begin{equation}
p \sim e^{-(m/m_P)^2},
\end{equation}
where $m_P$ is the Planck mass \cite{Christodoulou2016, Christodoulou2018}, consistently with a naive semiclassical expectation.   This probability is only significative when the black hole is Planckian, namely at the end of the Hawking evaporation, suggesting a black hole lifetime of order $m^3/\hbar$. But these calculations rely on a drastic truncation of the number of degrees of freedom, and it is possible that a more refined treatment could lower the black hole lifetime. In particular, arguments given in \cite{Haggard2014} suggest a black hole lifetime of the order of $m^2/\sqrt{\hbar}$. The question has important phenomenological consequences and at present is still open.

\section*{Acknowledgements}

We thank Fabio D'Ambrosio and Tommaso De Lorenzo for useful exchanges, and also Daniel Martinez for proofreading. CR thanks Jos\'e Senovilla and Ingemar Bengtsson for an interesting conversation and for pointing out the conical singularity in the firework metric. We acknowledge the OCEVU Labex (ANR-11-LABX-0060) and the A*MIDEX project (ANR-11-IDEX-0001-02) funded by the "Investissements d'Avenir" French government program managed by the ANR.

\vfill

\bibliographystyle{utcaps}
\bibliography{/home/pmd/Documents/Bibtex/B2W}

\end{document}